\documentclass[12pt]{article}
\usepackage{epsfig}

\def\beq{\begin{equation}}
\def\eeq#1{\label{#1}\end{equation}}
\def\eeqn{\end{equation}}

\def\beqa{\begin{eqnarray}}
\def\eeqa#1{\label{#1}\end{eqnarray}}
\def\eeqan{\end{eqnarray}}

\let\bar=\overbar

\def\Dslash{\not{\hbox{\kern-4pt $D$}}}
\def\dslash{\not{\hbox{\kern-2pt $\del$}}}

\def\msb{{\bar{\ssstyle M \kern -1pt S}}}

\def\Title#1{\begin{center} {\Large {\bf #1} } \end{center}}

\begin{document}

\Title{ $J/\psi$ production in association with \\ a charm-quark pair
at the Tevatron}

\bigskip\bigskip


\begin{raggedright}  

{\it Artoisenet Pierre\index{Artoisenet, P}\\
Center for Particle Physics and Phenomenology (CP3), \\
Universit\'e catholique de Louvain, B-1348 Louvain-la-Neuve, Belgium
}
\bigskip\bigskip
\end{raggedright}

\begin{abstract}
I study the direct hadroproduction of
$J/\psi$  associated with a charm-quark pair at leading order in 
$\alpha_S$ and $v$ in NRQCD.  This
process provides an interesting signature that could be studied at
the Tevatron.  
I consider both colour-singlet and colour-octet transitions.
I compare our results to the fragmentation approximation and discuss
the associated experimental signatures.

\end{abstract}

\section{Introduction}
Inclusive $J/\psi$ hadroproduction has been extensively studied 
in the theory of NRQCD~\cite{Bodwin:1994jh,yr}. A well-known 
innovation of this theory is the colour-octet mechanism: the
heavy-quark pair is allowed to be created in a colour-octet state over
short distances, the colour being neutralized over long distances.  It
is thanks to this very mechanism that it is possible to account for
the CDF data~\cite{Abe:1997jz,Abe:1997yz} on the inclusive $J/\psi$
and $\psi'$ cross sections at the
Tevatron~\cite{Braaten:1994vv,Cho:1995vh,Cho:1995ce}. 
However, the recent
data collected at $\sqrt{s}=1.96$ TeV by the CDF
collaboration~\cite{Abulencia:2007us} have revealed that the $J/\psi$
is unpolarised, in flagrant disagreement with the expectations of
NRQCD. For a recent review on $J/\psi$ production in hadron colliders, 
see \cite{Lansberg:2006dh}.

Another challenge to theorists has been provided by the recent
measurements at $B$ factories ($e^+e^-$ annihilation) \cite{Abe:2002rb}.
Surprisingly, it has been measured by the Belle collaboration
\cite{Uglov:2004xa} that the associated production $J/\psi +c \bar c$
almost saturates the inclusive production:
\begin{equation}
\frac{\sigma(e^+ e^- \rightarrow J/ \psi + c \bar{c})}{\sigma(e^+ e^- \rightarrow J/ \psi + X)}=0.59^{+0.15}_{-0.13}  \pm 0.12  \, .
\end{equation}
It is therefore natural to wonder whether the corresponding
production pattern could be large in hadroproduction as 
well~\cite{Artoisenet:2007xi}.   Besides
offering a new interesting signature, such as two leptons in
association with one or two charm-quark tags, this process contributes
to the $\alpha_S^4$ (NLO) corrections~\cite{Campbell:2007ws} to the inclusive 
hadroproduction of $J/ \psi$.

This note is organised as follows. In Section \ref{sec:LO_ggtoQqq},
we discuss the cross section and the polarisation of the $J/\psi$ produced 
with a $c\bar c $ pair, via colour-singlet and colour-octet transitions, 
at the Tevatron. In section \ref{sec:frag}, we compare our result to the fragmentation 
approximation.
In Section \ref{sec:asymptotic_behaviour}, we focus on the region of large transverse 
momentum and analyse the relevant experimental signatures in this regime.
Finally, Section \ref{sec:conclusions} is devoted to the conclusion of this note.

\section{The calculation of  $p\bar{p}\to J /\psi +  c\bar c $}
\label{sec:LO_ggtoQqq}

As for the case of open charm  cross sections, heavy-quarkonium 
hadroproduction is dominated by gluon fusion at Tevatron energy.  
We have checked that the light-quark initiated
process for $ J/\psi + c\bar{c}$ production is suppressed by three
orders of magnitude, and thus this contribution is neglected in the following.

In our numerical studies we have used:
\begin{itemize}
\item $\langle O_1(^3S_1) \rangle_{J/\psi}=1.16$ GeV$^3$; 
\item $\langle O_8(^3S_1) \rangle_{J/\psi}=1.06 \times 10^{-2}$ GeV$^3$ and 
$\langle O_8(^1S_0) \rangle_{J/\psi}=1 \times 10^{-2}$ GeV$^3$;
\item $\mu_0=\sqrt{(4 m_c)^2+P_T^2}$;
\item Br$(J/\psi \to \mu^+ \mu^- ) = 0.0588$; 
\item $m_c=1.5$ GeV;
\item pdf set:  CTEQ6M~\cite{Pumplin:2002vw}.
\end{itemize}

In Fig. 1 we show the $P_T$ distributions of $J/\psi$ 
 at the Tevatron, both for the colour-singlet and colour-octet
$J/\psi+c\bar c$ production.  
A cut on rapidity, $|y|<0.6$, selects centrally produced $J/\psi$'s.

For the colour-singlet case, we note that the $P_T$ distribution peaks
at $P_T \simeq m_c$ and then it starts a quick decrease,
dropping by four orders of magnitude at $P_T \simeq 20$ GeV.  
We verified that the topologies
where the $J/\psi$ is produced by two different quark lines 
always  dominate.
This behaviour is similar to the $J/\psi$ production at $e^+e^-$
colliders, where the presence of two extra charmed mesons seems to
indicate that the $J/\psi$ is mostly often created from two different
quark lines as well.  

Contrary to the situation in $e^+e^-$ collisions~\cite{Liu:2003jj}, 
the colour-octet production
plays an important role for the associated $J/\psi$ hadroproduction. Although 
negligible at small $P_T$, the $^3S_1^{[8]}$ transition starts to compete with the colour-singlet yield
at $P_T \approx 10 $ GeV, and dominates the associated production at larger values 
of the transverse momentum of the $J/\psi$. 
The integrated colour-octet cross section remains small compared to the colour-singlet one,
as a result of the  suppression of the long-distance matrix element for colour-octet
transitions.

\begin{figure*}[H]
\rotatebox{90}{\includegraphics[width=5cm]{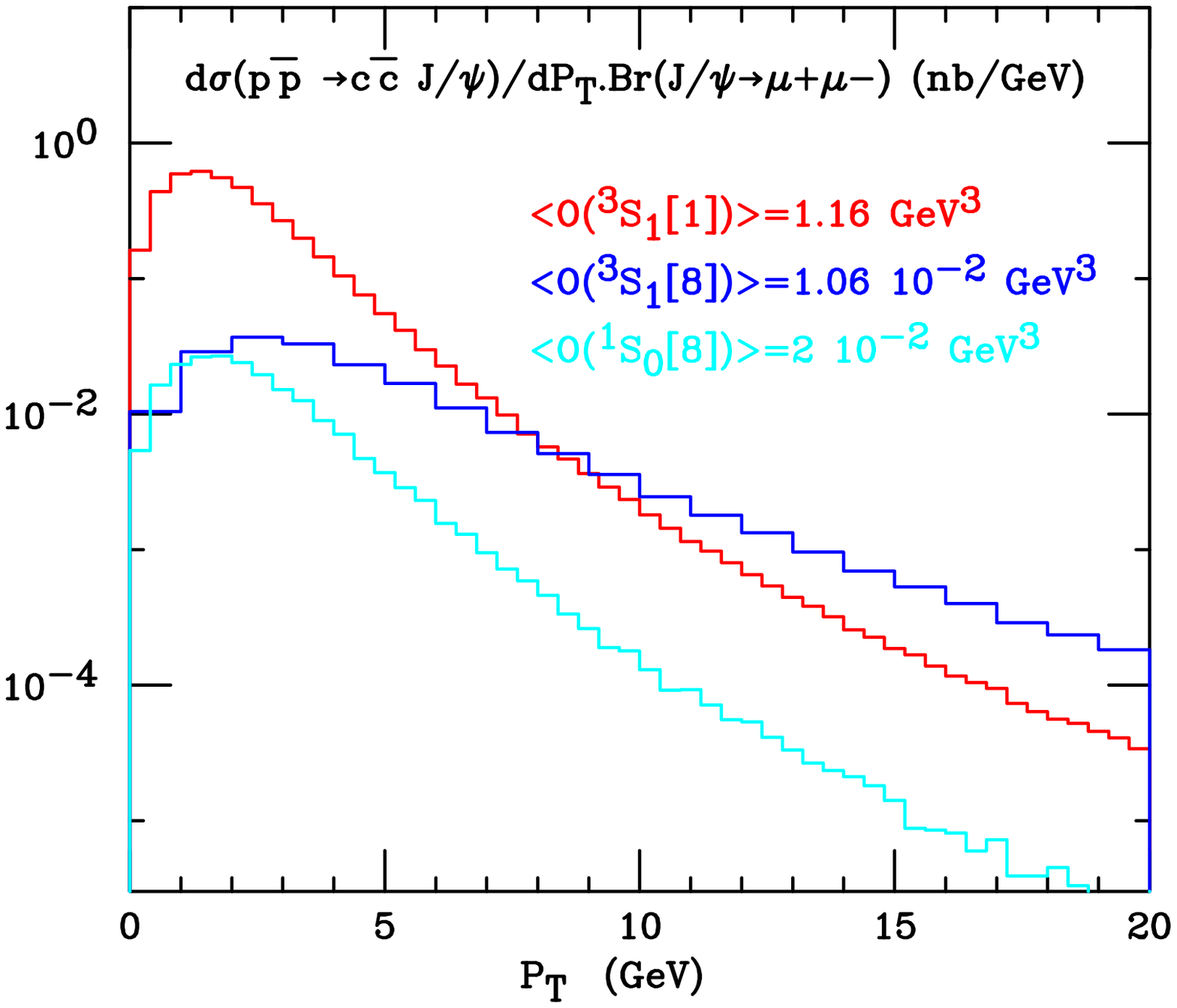}}
\quad
\rotatebox{90}{
\includegraphics[width=5cm]{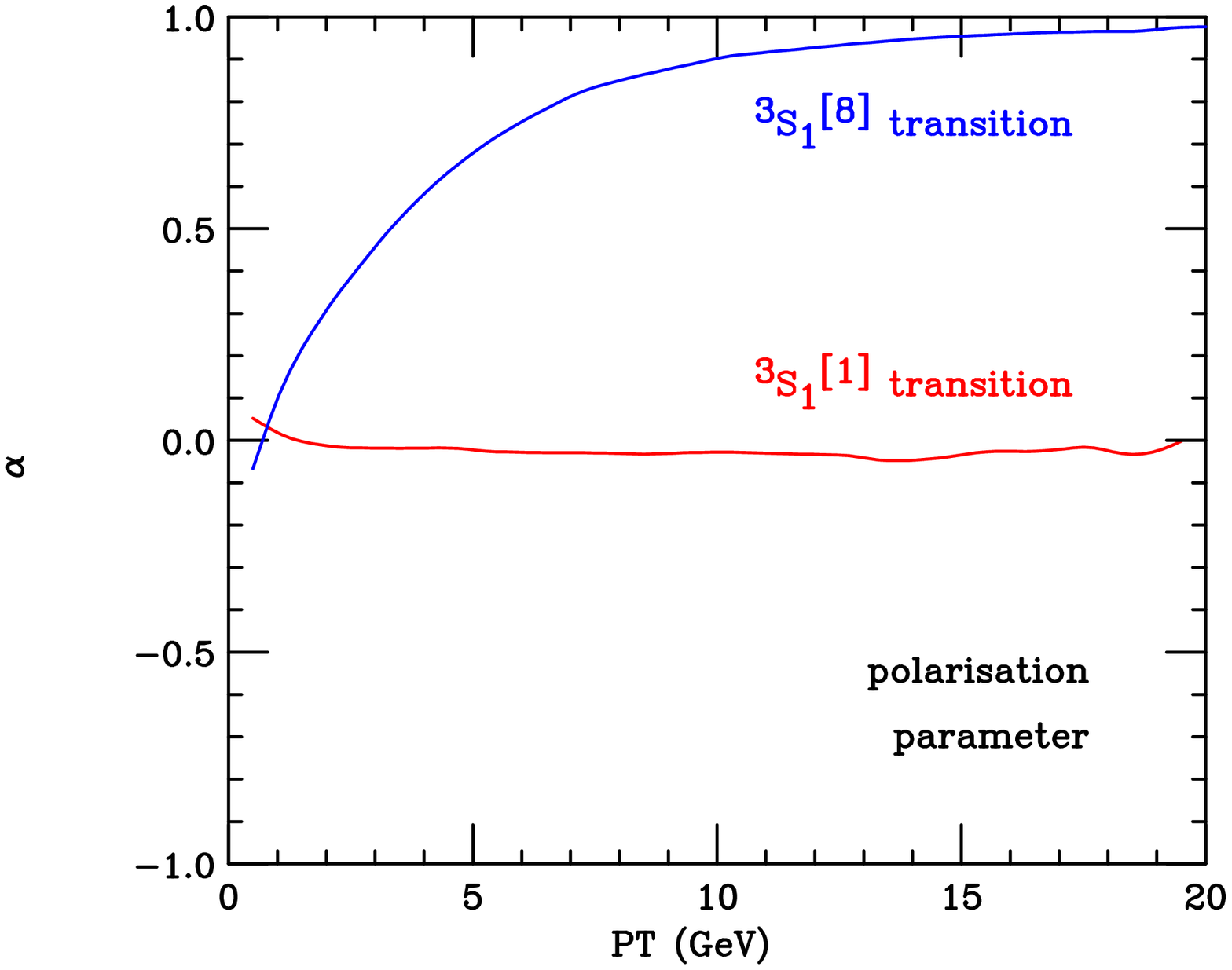}}
\caption{Left-hand-side: differential cross section for the process $p\bar{p}\to J/\psi +  c\bar c $  at the 
Tevatron, $\sqrt{s}=1.96$ TeV. 
Right-hand-side: polarisation parameter for the same process.
}
\label{fig:3S1QQ}
\end{figure*}

\begin{figure*}[H]
\quad
{\includegraphics[width=4cm]{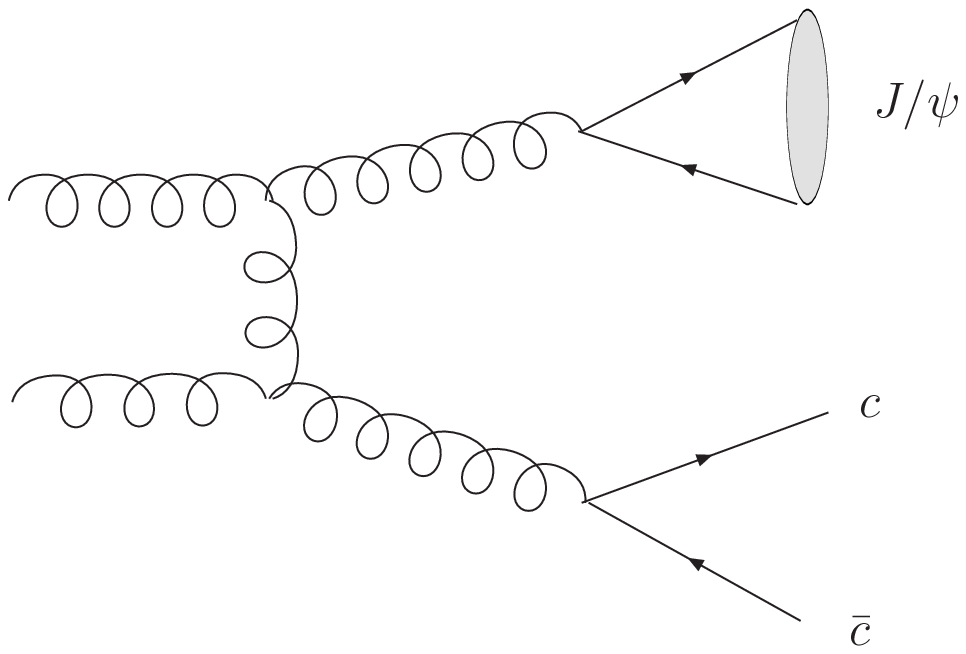} 
\quad \quad \quad  \quad
\includegraphics[width=5cm]{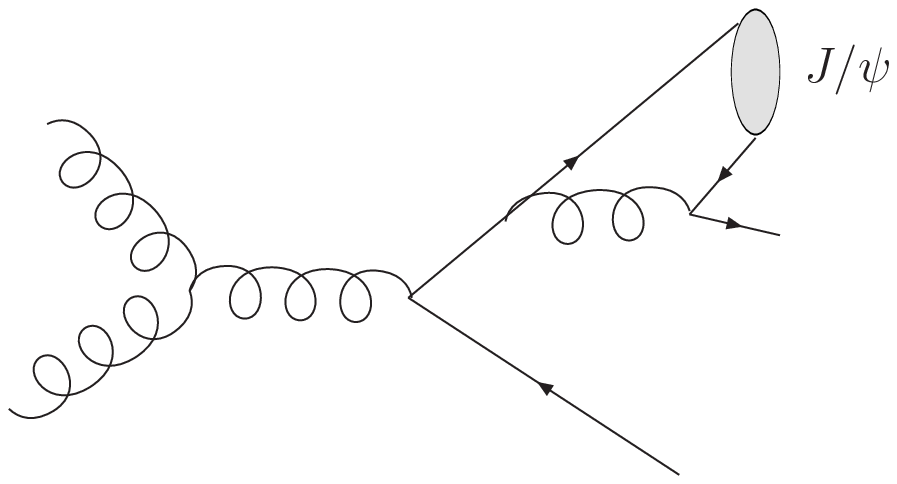}
} 
\caption{ Left-hand-side: typical Feynman diagram for the colour-octet $J/\psi$ 
production via the fragmentation of a gluon.  
Right-hand-side: typical Feynman diagram for the colour-singlet $J/\psi$ 
production via the fragmentation of a (anti-) charm quark.
}
\label{fig:frag_topo}
\end{figure*}

The preponderance of the $^3S_1^{[8]}$ transition
 at large $P_T$ is due to the kinematically-enhanced channel $gg \to c \bar c g \to c \bar c J/\psi$ 
where the $J/\psi$ is produced via the fragmentation of a 
nearly on-shell gluon (see left-hand-side of fig. \ref{fig:frag_topo}).
In this fragmentation topology, the partonic subprocess behaves like $\frac{1}{P_T^4}$.
The fragmentation of a gluon into a $J/\psi$ is not allowed in the case of the $\alpha_s^4$ 
colour-singlet production, which is therefore sub-dominant at large transverse momentum.
The colour-singlet fragmentation of a charm-quark into a $J/\psi$ 
(see right-hand-side of fig. \ref{fig:frag_topo})
 also behaves like $\frac{1}{P_T^4}$.
However, in this case, the invariant mass of the products of the fragmentation is larger than 
in the case of the gluon fragmentation. The charm-quark fragmentation is therefore expected to be 
kinematically suppressed compared to the gluon one, as it is confirmed by the $P_T$ 
spectrum in fig. \ref{fig:3S1QQ}.

The colour-octet $^1S_0$ transition has the same $P_T$ shape as the colour-singlet production,
but it is suppressed by one order of magnitude. Of course this suppression strongly depends 
on the values  chosen for the long distance matrix elements. 
Nevertheless, in the following, we will consider that the intermediate
state $^1S_0^{[8]}$ leads to a negligible cross section and will focus 
on the $^3S_1^{[1]}$
and $^3S_1^{[8]}$ transitions.

The polarisation of the quarkonium can  be determined
by analysing the angular distribution of the leptons.
Defining $\theta$ as the angle between one lepton 
direction in the quarkonium rest frame and the quarkonium direction in
the laboratory frame, the normalised angular distribution $I(\cos(\theta))$ is
\begin{equation}
 \label{angular_dist} I(\cos \theta) =
\frac{3}{2(\alpha+3)} (1+\alpha \, \cos^2 \theta)\,, 
\end{equation}
 where
the relation between $\alpha$ and the polarisation state of the
quarkonium is 
\begin{equation} 
\label{def_alpha}
\alpha=\frac{\sigma_T-2\sigma_L}{\sigma_T+2\sigma_L} \,. 
\end{equation}
The parameter $\alpha$ for colour-singlet and colour-octet associated 
productions  is displayed in fig. \ref{fig:3S1QQ}, right-hand-side. Whereas the $J/\psi$ produced via a colour-singlet transition is unpolarized independently of the value of its transverse momentum, the transverse polarisation state dominates the colour-octet $J/\psi$ production above $P_T=5$ GeV. This feature is to be related with the polarisation studies for the  $\alpha_s^3$ inclusive colour-octet $J/\psi$ production, which is also rapidly dominated by gluon fragmentation at large $P_T$.

\section{Testing the fragmentation approximation}
\label{sec:frag}

In the fragmentation approximation~\cite{Braaten:1993mp,Braaten:1994kd}, 
the cross section for the
production of a $J/\psi$ by gluon fusion via a parton
$i$ fragmentation is given, at all orders in $\alpha_S$, by 
\begin{equation}
d \sigma_{{J/\psi}}(P) =  \int^1_0 \, dz \, d\sigma \left(
gg \rightarrow i (\frac{P}{z},\mu_{frag}) + X \right)
D_{i\to{ J/\psi}}(z,\mu_{frag})
\end{equation}
 where $d\sigma \left(
gg \rightarrow i (\frac{P}{z},\mu_{frag}) + X\right)$ is the differential cross section to
produce an {\it on-shell} parton $i$ with
momentum $\frac{P}{z}$ and $D_{i\to{J/\psi}}(z,\mu_{frag})$ is the
fragmentation function of parton $i$ into a $J/\psi$.

The fragmentation scale, $\mu_{frag}$, is usually chosen to avoid
large logarithms of $P_T/\mu_{frag}$ in
$d \sigma$, that is $\mu_{frag}\simeq
P_T$. The resummation of the corresponding large logarithms of
$\mu_{frag}/m_Q$ appearing in the fragmentation function can be
 obtained via an evolution equation.

\begin{figure*}[H]
\rotatebox{90}{
\includegraphics[width=5cm]{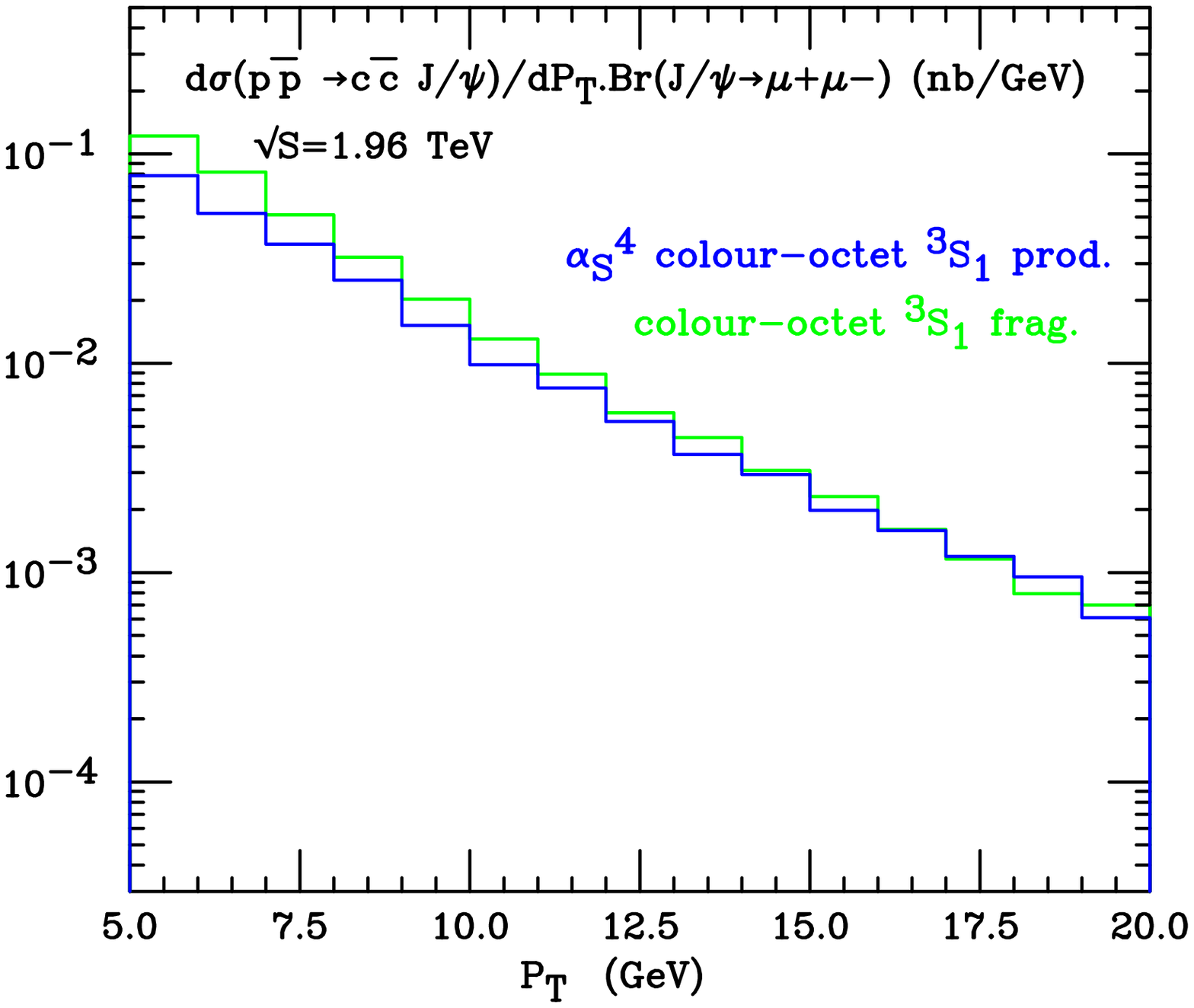}
}
\quad
\rotatebox{90}{\includegraphics[width=5cm]{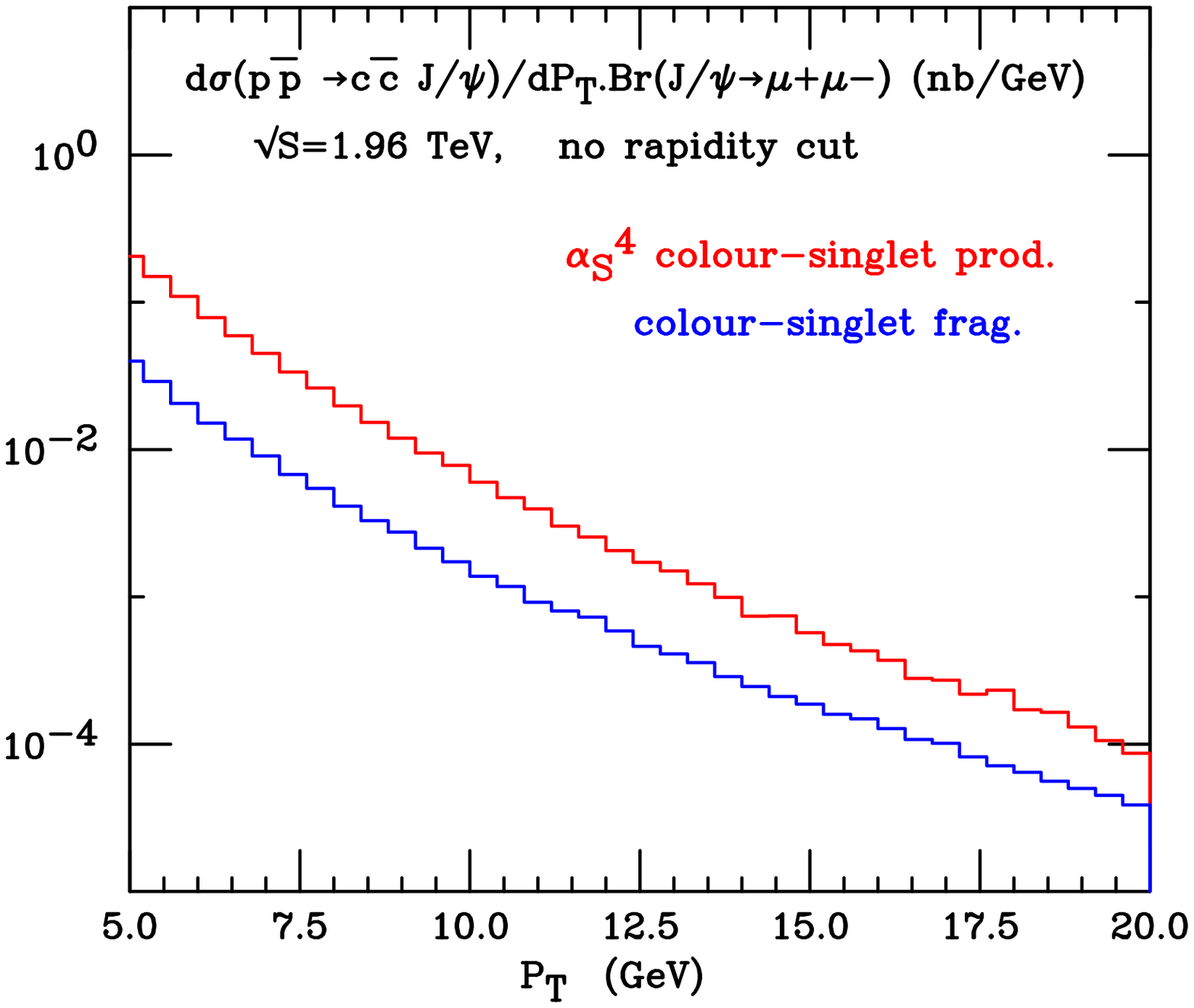}}
\caption{ Comparison between the  {\it full} LO cross section for $p\bar p\to J/\psi + c \bar c$
and the fragmentation approximation at $\sqrt{s}=1.96$ TeV, for  $^3S_1$ colour-octet transition (left-hand-side plot)  and  for colour-singlet transition (right-hand-side plot).
No cut on rapidity is applied. 
}
\label{fig:frag_approx}
\end{figure*}

We now turn to the  comparison of  our results 
for  the full LO cross sections
for $p\bar{p}\to J/\psi + c \bar c $ with those calculated in the
fragmentation approximation (without the evolution of the fragmentation
function). The same set of parameters of
Section~\ref{sec:LO_ggtoQqq} is employed\footnote{For consistency, the
coupling constant in the fragmentation function is
evaluated at the scale $\sqrt{(4 m_c)^2 +P_T^2}$. }.
Let us comment the colour-octet $^3S_1^{[8]}$  production first (fig. \ref{fig:frag_approx}, left-hand-side).
In this case, the fragmentation approximation overestimates the differential 
cross section by $50 \%$ at $P_T \approx 5$ GeV. However the gap between the two curves 
decreases rapidly so that the fragmentation approximation starts to be accurate 
within $10 \%$ above $P_T \approx 12$ GeV. In the case of the colour-singlet production,
the fragmentation  approximation is lower than the
full computation in the $P_T$ range accessible at the Tevatron.
We have verified that for $J/\psi$ production, the two curves still differ by a
little bit less than $10$\% at $P_T=80$ GeV.

So, contrary to the colour-octet gluon fragmentation, the  charm-quark fragmentation 
is not relyable in the kinematic region
accessible at the Tevatron. Again, this feature, as well as the predominance 
of the colour-octet yield at large $P_T$, is intuitively clear, since 
the virtual mass of the intermediate parton is smaller in the case of the 
colour-octet gluon fragmentation, it indeed remains fixed at the value $2m_c$
(and so the fragmentation function is proportional to $\delta(1-z)$). 

\section{Signatures at large transverse momentum}
\label{sec:asymptotic_behaviour}

\begin{figure*}[h]
\centerline{\mbox{
\rotatebox{90}{
\includegraphics[height=7cm]{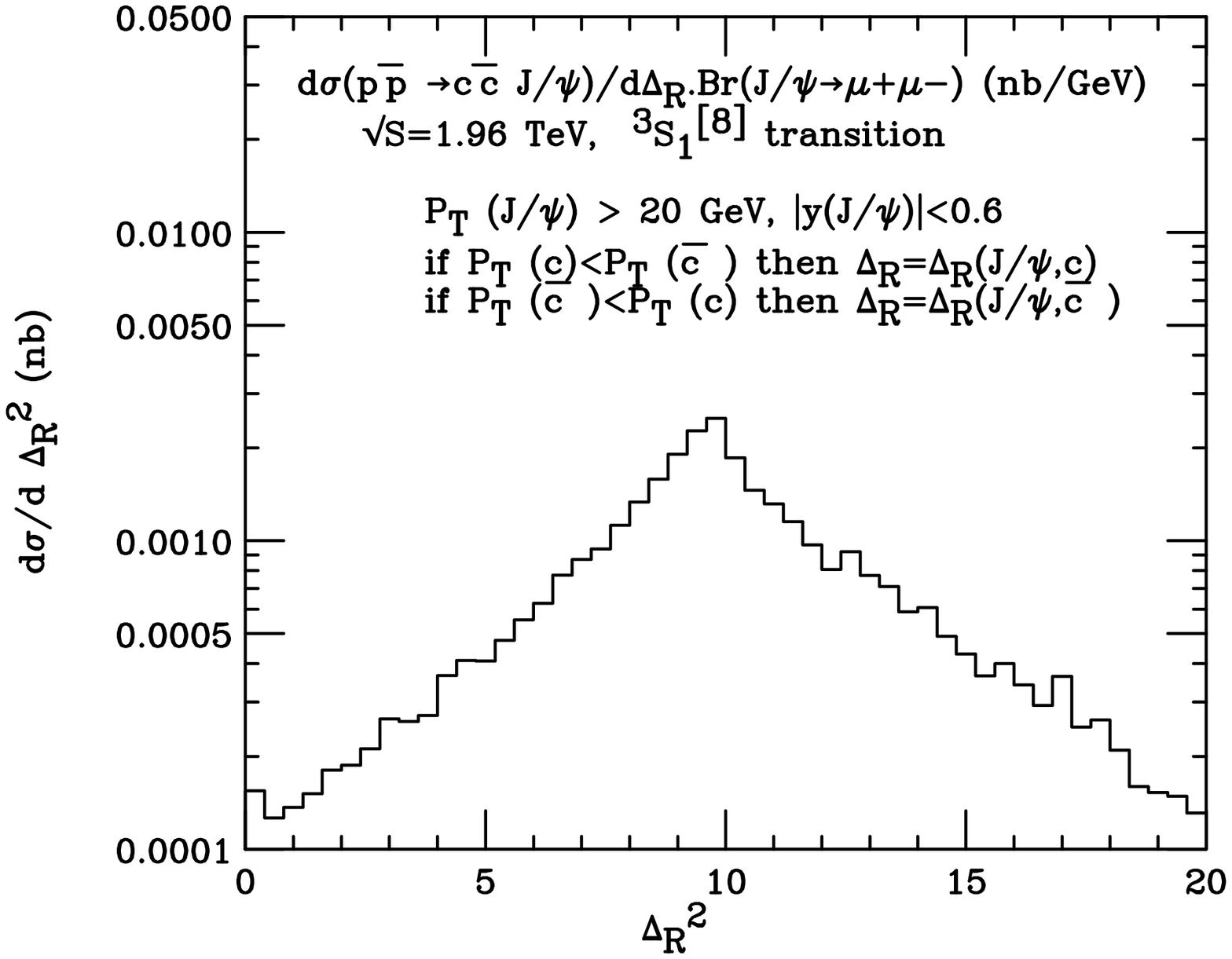}
}
\quad
\rotatebox{90}{
\includegraphics[height=7cm]{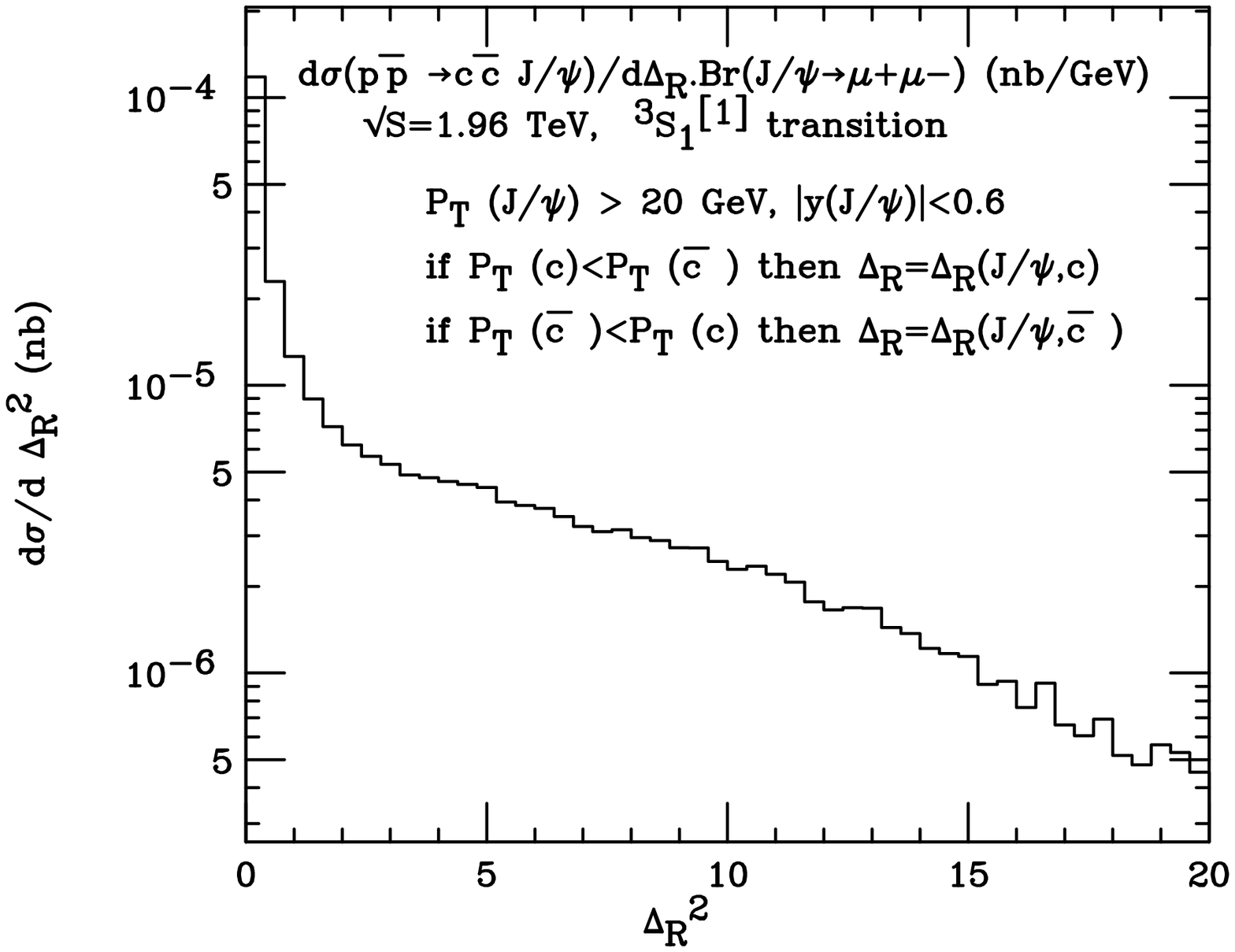}
}
}}
\caption{Angular separation between the $J/\psi$ and the $P_T$-softest charm quark, 
for the  $^3S_1^{[8]}$ transition (left-hand-side plot) and  for the $^3S_1^{[1]}$ transition 
(right-hand-side plot).}
\label{fig:deltaR}
\end{figure*}

Given that a charm-quark hadronizes most of the time into a heavy-light quark bound state,
the experimental signature of the process studied here is a $J/\psi$ accompanied with one or two charmed 
mesons. At large transverse momentum, both colour-singlet and colour-octet transitions contribute. 
From the study of the polarisation in section \ref{sec:LO_ggtoQqq}, we know that the angular distribution 
of the produced leptons can be used to discern these modes of production.
Another way to 
disentangle the two mechanisms is to study the distribution 
of events with respect to the angular separation between the 
$J/\psi$ and the $P_T$-softest charm quark.  We define 
$\Delta R^2=\Delta \phi^2+\Delta y^2$ to be the angular separation 
between the $P_T$-softest quark and the $J/\psi$. 
The resulting plots are displayed in Fig. \ref{fig:deltaR}.
 
For the colour-octet production, the $P_T$-softest charm-quark is most frequently emitted 
in the opposite direction of the $J/\psi$. This feature corroborates 
the validity of the fragmentation picture, where two gluons are emitted 
back-to-back, one of them giving rise to the $J/\psi$ via a colour-octet transition, the other emitting an open charm-quark pair.

For the colour-singlet production, we see in fig. \ref{fig:deltaR} 
that  the angular distribution is quite different. 
There is a peak at $\Delta R\approx 0$, corresponding to the situation where a 
charm-quark is emitted in the same direction as the $J/\psi$. 
This contribution is (approximately) taken into account in the 
fragmentation approximation, and is particularly interesting. Indeed,
it has been shown recently \cite{Nayak:2007mb}  that the 
cross section could be further enhanced in this region of the phase space, 
due to colour transfer between the unpaired charm-quark and one of the  active quarks. 

\begin{figure*}[h]
\centerline{\mbox{
\includegraphics[height=2cm]{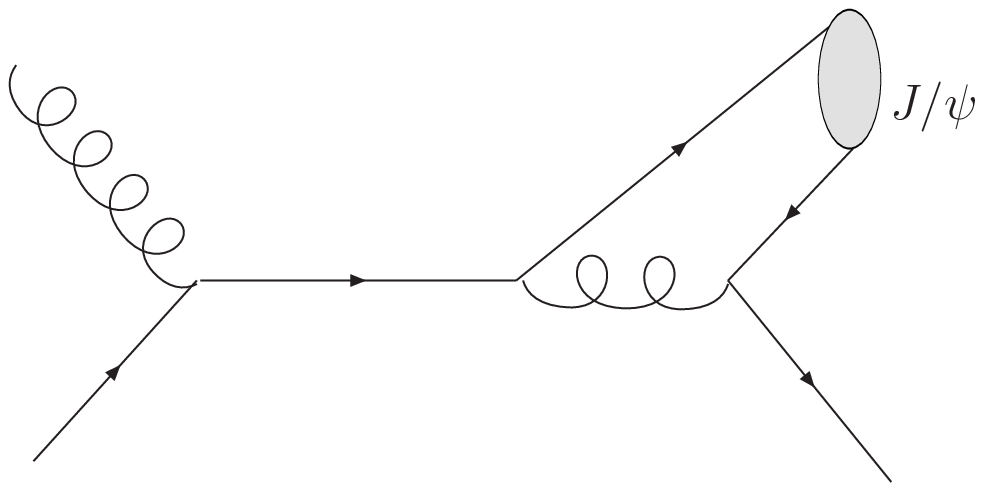}
\hspace{2 cm}
\includegraphics[height=2cm]{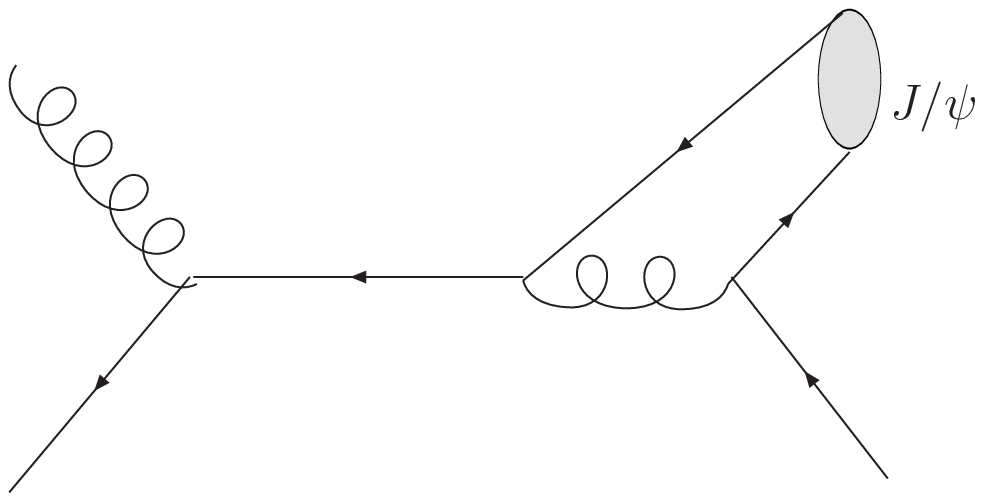}
}}
\caption{Leading order Feynman diagrams for the c-quark initiated $J/\psi$ production process.  }
\label{fig:cinitiated}
\end{figure*}

\begin{figure}[h]
\centerline{\mbox{
\rotatebox{90}{
\includegraphics[height=8cm]{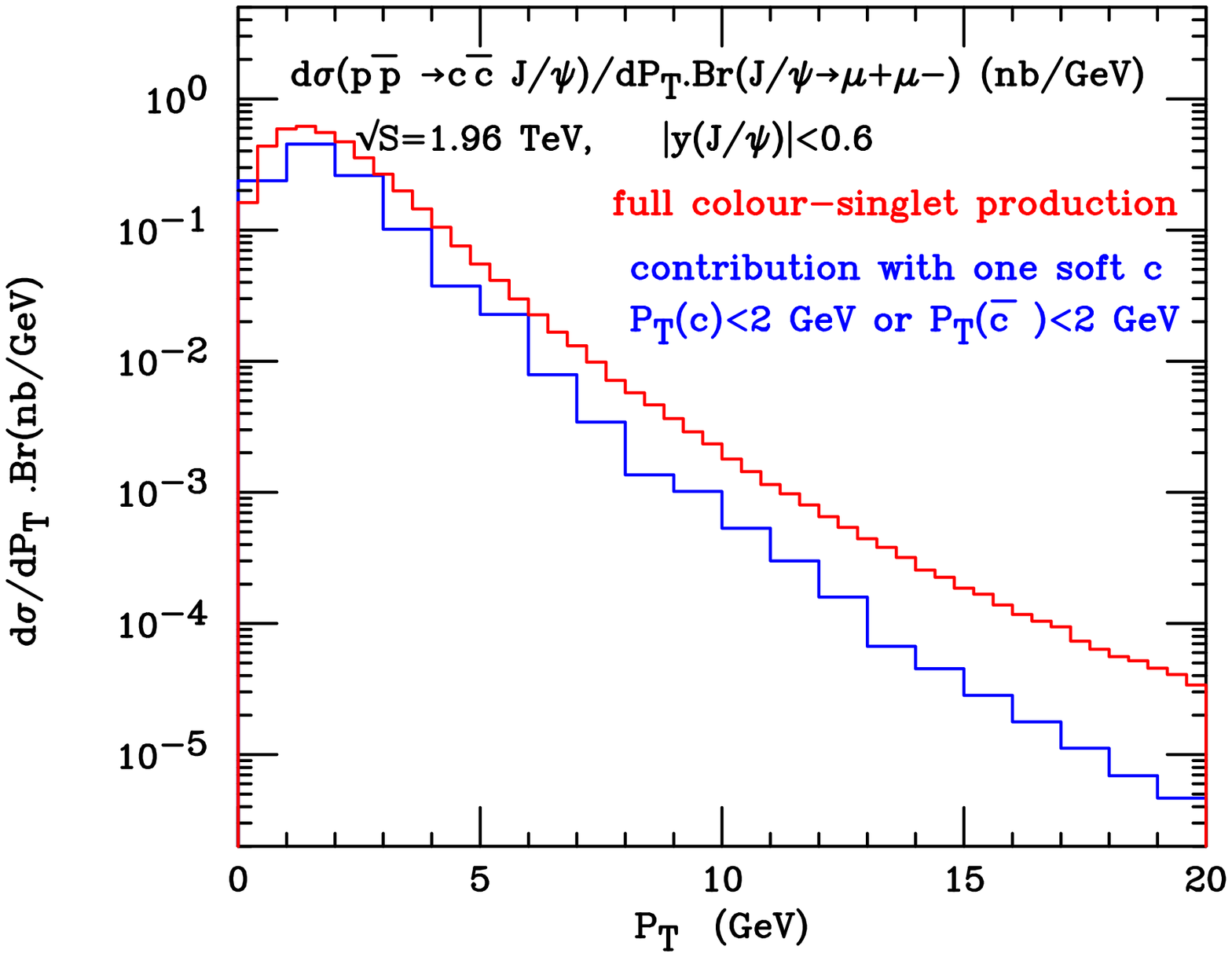}
}
}}
\caption{Red curve:  full colour-singlet LO associated production. Blue curve:  contribution from the phase space region where at least one open quark has a transverse momentum beneath $2$ GeV. }
\label{fig:softc_contrib}
\end{figure}

The rest of the distribution is pretty flat, indicating that, except
at $\Delta R \approx 0$, there is no 
privileged direction 
for the charm-quark. 
Contributions in this region have also  been partially considered through the study of the 
c-quark initiated process $cg \rightarrow c J/\psi$  (see Fig. \ref{fig:cinitiated}). 
Indeed, in this process, the initial $c$-quark originates from the splitting of 
a gluon into a pair of charm quarks collinear to the beam. 
One of the  quarks interacts with a gluon to produce the $J/\psi$, 
the other  quark is lost in the beam pipe. 
This configuration is included into the $\alpha_s^4$ process $gg \rightarrow c \bar c J/\psi$, in the phase space region where one of the open quarks is collinear to the beam.
A rough idea of this specific contribution can be obtained by requiring that one of the charm quark must have a small transverse momentum.
In Fig. \ref{fig:softc_contrib}, we plot the full colour-singlet production together with the curve resulting from the cuts
\begin{equation}
P_T(c)<2 \, \,\textrm{ GeV}, \quad \textrm{or} \quad P_T(\bar c )<2 \, \,\textrm{ GeV} .
\end{equation}

The curve associated to the presence of one soft-$P_T$ open quark becomes sub-dominant at large transverse momentum of the $J/\psi$. This is expected, since the presence of the two propagators in the $c$-quark initiated process (Fig. \ref{fig:cinitiated}) disfavours this mechanism at large $P_T$ compared to a fragmentation configuration.

\section{Conclusion}
\label{sec:conclusions}

In this note, we have presented the tree-level calculation for the
associated production of $J/ \psi$  with a
charm-quark pair. 
This process offers a new interesting signature,
that could be tested experimentally by measuring the fraction of
quarkonium produced with at least one heavy-light quark meson. 
Both colour-singlet and colour-octet transitions a priori contribute,
but can be disentangled by looking at the angular distribution of 
the leptons originating from the $J/\psi$, or by studying the
angular separation between  the $J/\psi$ and the charmed mesons.  

We have found that the charm-quark fragmentation approximation employed to
describe $J/\psi +c \bar c$ production should not be applied in the range of
transverse momenta reached at the Tevatron and analysed by the CDF
collaboration. This  approximation actually underestimates
the full colour-singlet production by more
than a factor four in the region $P_T \simeq 15$ GeV.  
On the other hand, the colour-octet gluon fragmentation is relevant 
in the region $P_T \geq 15$ GeV. 

In conclusion, we look forward to the measurement of the fraction of
events in the $J/\psi$ sample at the Tevatron, with at least one
charmed meson in the final state. Such a measurement could also
provide further insight to the mechanism responsible for inclusive
heavy-quarkonium production at hadron colliders.

\section*{Acknowledgments}
I would like to express my gratitude to 
my supervisor, Fabio Maltoni
for his advices and his training.
I thanks G. Bodwin, E. Braaten and J.P. Lansberg for usefull discussions.
The author is a Research Fellow of the \textit{Fonds
National de la Recherche Scientifique}, Belgium. 



\end{document}